\documentclass[twocolumn,showpacs,showkeys,preprintnumbers]{revtex4}
\usepackage{amssymb}

\usepackage{graphicx}
\usepackage{dcolumn}
\usepackage{bm}

\begin{document}

\preprint{APS/123-QED}
\title{Local Currents for a Deformed Algebra of Quantum Mechanics \\
with a Fundamental Length Scale\/}
\author{Gerald A. Goldin}
\email{gagoldin@dimacs.rutgers.edu}
\affiliation{\smallskip\\
Department of Physics, King's College London,\\
Strand, London WC2R 2LS, UK \smallskip \\
and \smallskip \\
Departments of Mathematics and Physics, Rutgers University,\\
118 Frelinghuysen Road, SERC 239, Busch Campus,\\
Piscataway, NJ 08854 USA }
\author{Sarben Sarkar}
\email{sarben.sarkar@kcl.ac.uk}
\affiliation{\smallskip\\
Department of Physics, King's College London,\\
Strand, London WC2R 2LS, UK \\
}
\date{\today}

\begin{abstract}
\noindent We explore some explicit representations of a certain
stable deformed algebra of quantum mechanics, considered by R.
Vilela Mendes, having a fundamental length scale. The relation of
the irreducible representations of the deformed algebra to those
of the (limiting) Heisenberg algebra is discussed, and we
construct the generalized harmonic oscillator Hamiltonian in this
framework. To obtain local currents for this algebra, we extend
the usual nonrelativistic local current algebra of vector fields
and the corresponding group of diffeomorphisms, modeling the
quantum configuration space as a commutative spatial manifold with
one additional dimension.
\end{abstract}

 \maketitle

--

\smallskip


\newtheorem{theorem}{Theorem} \newtheorem{acknowledgement}[theorem]{%
Acknowledgement} \newtheorem{algorithm}[theorem]{Algorithm}
\newtheorem{axiom%
}[theorem]{Axiom} \newtheorem{claim}[theorem]{Claim}
\newtheorem{conclusion}[%
theorem]{Conclusion} \newtheorem{condition}[theorem]{Condition}
\newtheorem{%
conjecture}[theorem]{Conjecture}
\newtheorem{corollary}[theorem]{Corollary}
\newtheorem{criterion}[theorem]{Criterion}
\newtheorem{definition}[theorem]{%
Definition} \newtheorem{example}[theorem]{Example}
\newtheorem{exercise}[%
theorem]{Exercise} \newtheorem{lemma}[theorem]{Lemma}
\newtheorem{notation}[%
theorem]{Notation} \newtheorem{problem}[theorem]{Problem} \newtheorem{%
proposition}[theorem]{Proposition}
\newtheorem{remark}[theorem]{Remark}
\newtheorem{solution}[theorem]{Solution}
\newtheorem{summary}[theorem]{%
Summary}



\section{Introduction}

\noindent The possibility of experimentally observing features of
quantum gravity at small length scales has heightened interest in
the study of spacetime noncommutativity
\cite{hss1947,cny1947}---e.g., through the mathematics of
noncommutative geometry \cite{ac1994}, and/or as a feature of
string theories \cite{mrd1998,fa1999,nem1999}. The characteristic
length scale $\ell$ at which nonclassical features of gravity
should emerge may be the Planck length $\,\ell_P = \sqrt{\hbar \,
G/c^3} \sim 1.6 \cdot 10^{-35}$ m, or it may be significantly
larger \cite{nah1998}.

One way to introduce such noncommutativity is algebraic. A few
years ago, Vilela Mendes \cite{vm2000} argued again for
consideration of the combined Heisenberg and Poincar\'{e} Lie
algebras as a kinematical algebra for relativistic quantum
mechanics. This structure is ``unstable'', but allows a
parameterized family of nontrivial deformations that are
``stable''---in the sense that all the Lie algebras in an open
neighborhood in the space of structure constants are mutually
isomorphic \cite{mg1964,an1967,mln1967,vm1994}. The nontrivial
second cohomology of the original Lie algebra is a necessary
condition for it to be deformable \cite{da1995}. The proposed
stable algebra for relativistic quantum mechanics is a deformation
by two parameters $\ell$ and $R$, which are fundamental lengths.
Taking $\ell \rightarrow 0$ and $R\rightarrow \infty$ leads to
recovery of the original Lie algebra. Vilela Mendes argues on
fundamental grounds for describing the physical world by means of
a stable Lie algebra, so that small changes in physical constants
do not fundamentally alter the structure. We take no position here
on this question, but explore some interesting consequences of the
Vilela Mendes approach. A recent, beautiful paper of
Chryssomalakos and Okon describes and discusses the full set of
possible stable deformations of the Heisenberg-Poincar\'e algebra,
with explanation of the relevant cohomology theory and detailed
references \cite{cceo2004}.

The present article is motivated by the problem of defining an
equal-time, local current algebra compatible with the
nonrelativistic quantum kinematics that follows from Vilela
Mendes' proposal. Like him we consider the case where
$R\rightarrow \infty $ but $\ell \neq 0$; then the space-time
coordinate operators no longer commute. We clarify the relation of
the irreducible representations of a deformed subalgebra to those
of the limiting Heisenberg algebra, concentrating on the case of
one space dimension (although our considerations generalize
straightforwardly to higher dimensions). The limit procedure here
goes back to a 1970 scheme of Barut and Bohm for reduction of
certain representations of $SO(4,2)$ \cite{bb1970}. But our
construction of the generalized kinetic energy and harmonic
oscillator Hamiltonians in this framework leads to an answer
different from that suggested by Vilela Mendes.

One way of obtaining local currents for the deformed algebra is to
extend the usual nonrelativistic local current algebra (LCA) of
scalar functions and vector fields, and the corresponding
infinite-dimensional groups of scalar functions and
diffeomorphisms. In doing this we make use of an abstract
single-particle configuration space, which is a commutative
spatial manifold having one dimension more than the configuration
space for the limiting situation with $\ell \rightarrow 0$. Thus
the deformed $(1+1)$-dimensional theory entails self-adjoint
representations of an infinite-dimensional Lie algebra of
nonrelativistic, local currents for a (2+1)-dimensional space-time
(LCA2). To be able to recover the usual current algebra (LCA1) in
the limit $\ell \rightarrow 0$, one may introduce a semidirect sum
of LCA2 with the algebra of vector fields of the line. The local
operators then act in a direct integral of irreducible
representations of the global, finite-dimensional deformed Lie
algebra. This seems to open up interesting new possibilities,
which we discuss briefly.

The paper is organized as follows. In Sec. $\!$II, we present the
necessary background---the $(\ell, R)$-deformed Lie algebra of
relativistic quantum mechanics, the nonrelativistic local current
algebra and its relation to Heisenberg algebra, and the desirable
properties for local currents in relation to the deformed algebra
with $\ell \neq 0$. In Sec. $\!$III, we review several different,
but unitarily equivalent, self-adjoint representations of the
$\ell$-deformed Heisenberg algebra for the case of one spatial
dimension---which is isomorphic to the Lie algebra of the group of
rigid motions of the plane. This subalgebra for the
one-dimensional problem serves as a useful laboratory. We write
explicitly the unitary operators intertwining the representations.
This permits clarification of how an irreducible representation of
the usual Heisenberg algebra can be recovered in the $\ell \to 0$
limit. In Sec. $\!$IV, we discuss the kinetic energy and harmonic
oscillator Hamiltonians. In Sec. \negthinspace V, we consider
first the problems associated with currents localized with respect
to the spectrum of the deformed position operator. Then we develop
and discuss the extended nonrelativistic local current algebra and
diffeomorphism group.

\section{Background}

\noindent Introducing the $4$-vectors $q_{\mu }$ and $p_{\nu }$,
$\mu ,\nu =0,1,2,3$, and the Lorentz generators $M_{\mu \nu }$,
one combines the canonical brackets,
\begin{eqnarray}
\left[\,p_{\mu },q_{\nu }\right] &=&\,\,i\,\hbar \,\eta _{\mu \nu
}\,\mathcal{J},
\label{heisenberg} \\
\left[\,q_{\mu },q_{\nu }\right] \,=\, \left[ p_{\mu },p_{\nu
}\right] &=& \left[\,q_\mu, \mathcal{J}\,\right] \,=\,
\left[\,p_\mu, \mathcal{J}\,\,\right] \,=\,0\,, \nonumber
\end{eqnarray}
with the Lorentz brackets,
\begin{eqnarray}
\left[ M_{\mu \nu },M_{\rho \sigma }\right] &=&  \nonumber \\
i(M_{\mu \sigma }\eta _{\nu \rho }+M_{\nu \rho }\eta _{\mu \sigma
} &-&M_{\nu \sigma }\eta _{\mu \rho }-M_{\mu \rho }\eta _{\nu
\sigma }), \label{lorentz}
\end{eqnarray}%
by means of the additional brackets,
\begin{eqnarray}
\left[ M_{\mu \nu },p_{\lambda }\right] &=&i\left( p_{\mu }\eta
_{\nu
\lambda }-p_{\nu }\eta _{\mu \lambda }\right) \,,  \nonumber \\
\left[ M_{\mu \nu },q_{\lambda }\right] &=&i\left( q_{\mu }\eta
_{\nu \lambda }-q_{\nu }\eta _{\mu \lambda }\right) \,,
\label{covariance}
\\ \left[M_{\mu \nu}, \mathcal{J}\,\,\right] &=& 0\,, \nonumber
\end{eqnarray}%
where $\eta _{\mu \nu }=$ diag$\,\left[ 1,-1,-1,-1\right]$ in
units with $c = 1$. To describe the quantum kinematics of a
particle, one typically represents a {\it subalgebra\/} of this
Lie algebra by self-adjoint operators in Hilbert space.

While the Lie algebras of Eqs. \negthinspace (\ref{heisenberg})
and Eqs. \negthinspace (\ref{lorentz}) are separately stable, the
combined Lie algebra of Eqs. \negthinspace
(\ref{heisenberg})-(\ref{covariance}) is not; so we consider a
stable deformation.

\subsection{Deformed Lie algebras for quantum mechanics}

The relevant deformation is labeled by fundamental lengths $R$ and
$\ell $, and satisfies brackets where Eqs. \negthinspace
(\ref{lorentz}) and
(\ref%
{covariance}) are unchanged, but Eqs. \negthinspace
(\ref{heisenberg}) are replaced by,
\begin{eqnarray}
\left[\,p_{\mu },q_{\nu }\right] &=&\,\,i\,\hbar \,\eta _{\mu
\nu}\,\mathcal{J}
\nonumber \\
\left[\,q_{\mu },q_{\nu }\right] &=&-i\varepsilon \,\ell
^{\,2}\,M_{\mu \nu
}\,, \nonumber \\
\left[\,p_{\mu },p_{\nu }\right] &=&-i\frac{\varepsilon ^{\,\prime
}\hbar ^{\,2}}{R^{2}}\,M_{\mu \nu }\,, \label{generalized}
\\
\left[\,q_\mu, \mathcal{J}\,\right] &=& i\varepsilon \frac{\ell
^{2}}{\hbar }%
p_{\mu } \,, \nonumber \\
\left[\,p_\mu, \mathcal{J}\,\right] &=&
\,-i\frac{\varepsilon^{\,\prime}\hbar}{R^2}\,q_\mu\,, \nonumber
\end{eqnarray}%
where $\varepsilon $ and $\varepsilon ^{\prime }$ are $\pm 1$.
Evidently as $%
\ell \rightarrow 0$ and $R\rightarrow \infty $, we recover Eqs.
\negthinspace (\ref{heisenberg})-(\ref{covariance}).

This Lie algebra is isomorphic to the Lie algebra of the
orthogonal group in six dimensions, with metric $\eta _{ab}=$
diag$\,\left[ 1,-1,-1,-1,\,\varepsilon ^{\,\prime },\,\varepsilon
\,\right] $. Evidently in a self-adjoint representation, the
$\,q_{\mu }\,$ no longer commute with each other. Their
interpretation as space-time coordinate operators in such a
representation may be questioned \cite{cceo2004}; but if we
maintain this interpretation, Heisenberg-like uncertainty
relations for these coordinates suggest that space-time becomes
``fuzzy'' to order $\ell$ \cite{ec2001}.

We remark that in other specific models, the noncommutativity of
the coordinate operators is different. For example, in the case of
a charged particle moving in a plane perpendicular to a magnetic
field of magnitude $B$, in the limit as the mass $\mathfrak{m} \to
0$ we expect, for $j,k = 1,2$,
\begin{equation}
\left[\,q_{j}, q_{k}\right] \,=\, i\,\theta _{jk}
\mathcal{J}\,,\quad \left[\,q_j, \mathcal{J}\,\right] = 0\,,
\label{standard}
\end{equation}
where $\theta$ is a constant antisymmetric matrix inversely
proportional to $B$. A similar bracket occurs for a bosonic string
when there is a background, constant Neveu-Schwarz $2$-form in the
world volume of a D-brane \cite{mrd2001, szabo2001}.

As the parameter $\ell$ is relevant locally, we shall follow
Vilela Mendes in focusing on the algebra obtained by taking $R
\rightarrow \infty $. Then the brackets involving $R$ in Eqs.
\negthinspace(\ref{generalized}) become zero. We now want to
concentrate on self-adjoint representations of the Heisenberg-like
subalgebra, with $j,k = 1,2,3$, and with $\varepsilon =-1$, given
by the spatial components of Eqs.
\negthinspace(\ref{lorentz})-(\ref{covariance}), together with the
brackets,
\begin{eqnarray}
\left[\,q_{j},q_{k}\right] &=&i\,\ell^{2}M_{jk}\,,  \nonumber \\
\left[\,q_{j},p_{k}\right] &=&i\delta_{jk}\hbar \mathcal{J}\,,
\nonumber \\
\left[\,q_{j},\mathcal{J}\,\right] &=&-i\frac{\ell ^{2}}{\hbar
}p_{j}\,, \label{algebra} \\
\left[\,p_{j},p_{k}\right] &=& \left[\,p_{j},\mathcal{J}\,\right]
\,=\,0\,. \nonumber
\end{eqnarray}
The Lie algebra of Eqs.
\negthinspace(\ref{lorentz})-(\ref{covariance}) and
(\ref{algebra}) represents the {\it global\/} symmetry of the
deformed quantum theory. We desire, however, to incorporate a
description of {\it local\/} symmetry, for which we turn to local
current algebra.

\subsection{Equal-time local current algebra}

In standard, nonrelativistic quantum theory, the second-quantized
field $\widehat{\psi }\left( \mathbf{x},t\right)$ and its adjoint
$\widehat{\psi }^*\left( \mathbf{x},t\right)$, for $\mathbf{x} \in
\mathbb{R}^d$, are operator-valued distributions in Fock space.
These satisfy the equal-time canonical commutation $(-)$ or
anticommutation $(+)$ relations,
\begin{equation}
\left[\,\widehat{\psi }\left( \mathbf{x}\right),
\widehat{\psi}^*\left(\mathbf{y}\right) \right]_{\pm \,
(\mathrm{fixed}\,t)} \,=\, \delta(\mathbf{x} - \mathbf{y})\,;
\end{equation}
the argument $t$ is henceforth suppressed. Then the local,
fixed-time mass density $\rho \left( \mathbf{x}\right)$ and
momentum density $J\left( \mathbf{x}\right)$ are operator-valued
distributions, defined formally by
\begin{eqnarray}
\rho \left( \mathbf{x}\right) &=& \mathfrak{m}\widehat{\psi
}^{\ast }\left( \mathbf{x}
\right) \widehat{\psi }\left( \mathbf{x}\right)\,, \nonumber \\
\mathbf{J}\left( \mathbf{x}\right) &=& \frac{\hbar }{2i}\left\{
\widehat{\psi } ^{\ast }\left( \mathbf{x}\right) \nabla
\widehat{\psi }\left( \mathbf{x} \right) -\left[ \nabla
\widehat{\psi }^{\ast }\left( \mathbf{x}\right) \right]
\widehat{\psi }\left( \mathbf{x}\right) \right\}.\quad
\end{eqnarray}
These obey a certain singular Lie algebra, which is independent of
whether the original field is bosonic or fermionic \cite{rd1968}.
Define
\begin{eqnarray}
\rho \left( f\right) &=& \int \rho \left( \mathbf{x}\right)
f\left( \mathbf{x} \right) d\mathbf{x}\,,  \nonumber \\
J\left( \mathbf{g}\right) &=& \int \sum_{k=1}^d J_{k}\left(
\mathbf{x}\right) g_{k}\left( \mathbf{x}\right) d\mathbf{x}\,,
\label{current}
\end{eqnarray}
where $f$ and the components $g_k$ of $\mathbf{g}$ are
compactly-supported $C^\infty$ test functions on $\mathbb{R}^d$.
Then one obtains the local current algebra \cite{gag1971},
\begin{eqnarray}
\left[ \rho \left( f_{1}\right) ,\rho \left( f_{2}\right) \right]
&=&0\,,
\nonumber \\
\left[ \rho \left( f\right) ,J\left( \mathbf{g}\right) \right]
&=&i\hbar
\rho \left( \mathbf{g}\cdot\nabla f\right)\,,  \label{current2} \\
\left[ J\left( \mathbf{g}_{1}\right) ,J\left(
\mathbf{g}_{2}\right) \right] &=&-i\hbar J\left( \left[
\mathbf{g}_{1},\mathbf{g}_{2}\right] \right)\,; \nonumber
\end{eqnarray}
where $\,\left[\,\mathbf{g}_{1},\mathbf{g}_{2}\right] =
\mathbf{g}_{1} \cdot \nabla \mathbf{g}_{2}-\mathbf{g}_{2} \cdot
\nabla \mathbf{g}_{1}$ is the usual Lie bracket of vector fields.

In the $1$-particle Hilbert space $L_{d{\mathbf
x}}^{2}\left({\mathbb R}^{d}\right)$, we have the self-adjoint
representation
\begin{eqnarray}
\rho \left( f\right)\Psi({\mathbf x}) &=& \mathfrak{m} f({\mathbf
x})\Psi({\mathbf x})\,,
\label{rep} \\
J\left( \mathbf{g}\right)\Psi({\mathbf x}) &=& \frac{\hbar
}{2i}\{\,\mathbf{g} \left( \mathbf{x}\right) \cdot \mathbf{\nabla
}\Psi({\mathbf x}) + \mathbf{\nabla } \cdot [\mathbf{g}\left(
\mathbf{x}\right)\Psi({\mathbf x}) ]\}\,, \nonumber
\end{eqnarray}
where $\mathfrak{m}$ is the particle mass. Now as the test
function $f\left( \mathbf{x}\right)$ approaches an indicator
function $\chi _{B}\left( \mathbf{x}\right)$ for a Borel set $B
\subseteq \mathbb R^{d}$, the expectation value $(\Psi,\rho \left(
f\right)\Psi)$ with respect to the single-particle wave function
$\Psi$  approximates $\,\mathfrak{m}\int \chi _{B}\left(
\mathbf{x%
}\right) \left| \psi \left( \mathbf{x}\right) \right|
^{2}d\mathbf{x}$, which is the mass times the usual probability
for finding the particle in the region $B$. If $f\left(
\mathbf{x}\right)$ approaches $\delta \left( \mathbf{x-x}%
_{0}\right) $ for a fixed point ${\mathbf x}_{0} \in
\mathbb{R}^d$, then $(\Psi,\rho \left( f\right)\Psi)$ approaches $
\mathfrak{m}\left| \Psi \left( \mathbf{x}_{0}\right) \right|
^{2}$. We also see how the Heisenberg algebra is recovered---if
$f\left(\mathbf{ x}\right)$ approximates the coordinate function
$x_{j}$, then $\rho(f)$ approximates the moment operator
$\,\mathfrak{m}q_{j}\,$ acting in $L^2_{d{\mathbf
x}}(\mathbb{R}^d)$ {\it via\/} multiplication by
$\,\mathfrak{m}x_j$. Similarly, if ${\mathbf g}({\mathbf x})$ is
taken to approximate a constant vector field in the $j$-direction,
so that (let us say) $ g_{j}\left( \mathbf{x}\right) \sim 1$ with
$g_{k}\left( \mathbf{x}\right) =0$ for $k\neq j$, then $J\left(
\mathbf{g}\right) \sim -i\hbar
\partial / \partial x_{j}$, which is the action of the momentum
operator $\,p_{j}\,$ in $\,L^2_{d{\mathbf x}}(\mathbb{R}^d)$.

In short, the LCA in the $1$-particle representation, with
suitable (global) choices of test functions, allows recovery of
the usual quantum-mechanical representation of the
finite-dimensional subalgebra of Eqs.
\negthinspace(\ref{heisenberg}) having spatial indices. Likewise,
generators of spatial rotations may be recovered---e.g., in the
$1$-particle representation in three space dimensions, the
operator for orbital angular momentum about the $x_3$-axis is
approximated by choosing $\,g_{1}\left( \mathbf{x}\right) =
-x_{2},\,g_{2}\left( \mathbf{x} \right) = x_{1}$, and
$\,g_{3}\left( \mathbf{x}\right) = 0\,$ inside a large compact
region $\,\left| \mathbf{x}\right| \leq R$; outside this region,
$\,\mathbf{g}\left( \mathbf{x}\right) $ falls smoothly to $0$.
Then $J\left( \mathbf{g}\right)$ approximates the operator
$\,\hbar M_{12} = L_3 = (\mathbf{q} \times \mathbf{p}) \cdot
\mathbf{e}_{3}$ acting in $\,L^2_{d{\mathbf x}}(\mathbb{R}^3)$,
where $\mathbf{e}_{3}$ is the unit vector in the
$\,x_3$-direction.

The study of inequivalent, self-adjoint representations of this
infinite-dimensional algebra has turned out to be a powerful
method for classifying, and in some cases predicting, kinematical
possibilities for quantum systems. These possibilities include the
usual $N$-particle representations, $N = 1,2,3,\dots$, satisfying
bosonic or fermionic statistics for $N \geq 2$ in more than one
space dimension. They also include particle systems obeying
anyonic statistics in two-dimensional space \cite{jml1977,
gag1980, gag1981, fw1982, gag1983a} or other exotic statistics,
particles with spin \cite{gag1983b, gag1983c}, composite systems
having dipole or higher multipole moments \cite{gag1985}, and
infinite-particle or extended systems having infinite-dimensional
configuration spaces \cite{gag1974, sa1999, gag2004}.

This is our motivation for investigating the possibilities for
defining local currents appropriate to the deformed Lie algebra of
Eqs. \negthinspace(\ref{algebra}). The goal is to obtain an
infinite-dimensional, local Lie algebra that is a deformation or
extension of the LCA of Eqs. \negthinspace(\ref{current2}), which,
with suitable choices of test functions in a $1$-particle
representation, allows recovery of a standard representation of
Eqs. \negthinspace(\ref{algebra}). Then the unitarily inequivalent
representations of the deformed LCA should describe kinematical
possibilities for a nonrelativistic version of the deformed
quantum theory. Thus in Sec. \negthinspace III, we shall discuss
some different ways of writing standard representations of Eqs.
\negthinspace(\ref{current2}).

Note that we may introduce an operator-valued distribution
$Q(f,{\bf g})$ acting in $L_{d{\mathbf x}}^{2}\left({\mathbb
R}^{d}\right)$, defined by
\begin{eqnarray}
Q(f,{\mathbf g})\Psi \,=\, f({\bf x})\Psi \,\,+& & \nonumber \\
\frac{1}{2i}\,\{\,\mathbf{g} \left( \mathbf{x}\right) \cdot
\mathbf{\nabla }\Psi({\mathbf x}) &+& \mathbf{\nabla } \cdot
[\mathbf{g}\left( \mathbf{x}\right)\Psi({\mathbf x}) ]\}\,
\label{semidirectsum1}.
\end{eqnarray}
Then $Q$ is a self-adjoint representation of the natural
semidirect sum of the commutative Lie algebra of
compactly-supported, real-valued $C^\infty$ functions $f$ on
$\mathbb{R}^d$, with the Lie algebra of vector fields ${\mathbf
g}$ on $\mathbb{R}^d$; namely,
\begin{equation}
[(f_1,\mathbf{g}_1), (f_2,\mathbf{g}_2)] = (\mathbf{g}_2\cdot
\nabla f_1 - \mathbf{g}_1 \cdot \nabla f_2,
-[\mathbf{g}_1,\mathbf{g}_2])\,.\, \label{semidirectsum2}
\end{equation}
The physical constants $\mathfrak{m}$ and $\hbar$ do not enter
(\ref{semidirectsum1}), but Eqs. \negthinspace(\ref{rep}) follow
from it when we set $\,\rho(f) = \mathfrak{m}\,Q(f,0)\,$ and
$J(\mathbf{g}) = \hbar \,Q(0,\mathbf{g})$.

The group that is associated with Eq.
\negthinspace(\ref{semidirectsum2}) is the natural semidirect
product ${\mathcal D}({\mathbb R}^d) \times ${\it
Diff\/}$^c({\mathbb R}^d)$, where $\mathcal{D}({\mathbb R}^d)$ is
the group of compactly supported, real-valued $C^\infty$ functions
on $\mathbb{R}^d$ under pointwise addition, and {\it
Diff\/}$^c({\mathbb R}^d)$ is the group of compactly supported
diffeomorphisms of ${\mathbb R}^d$ under composition
\cite{gag2004}. These groups are endowed with the topology of
uniform convergence in all derivatives on compact sets. The group
law, for $f_1, f_2 \in {\mathcal D}({\mathbb R}^d)$ and $\phi_1,
\phi_2 \in$ {\it Diff\/}$^c({\mathbb R}^d)$, is
\begin{equation}
\left( f_{1},\phi _{1}\right) \left( f_{2},\phi _{2}\right)
=\left( f_{1}+f_{2}\circ \phi _{1},\phi _{2}\circ \phi
_{1}\right)\,, \label{group1}
\end{equation}
where $\circ$ denotes composition.

Given the compactly-supported, $C^\infty$ vector field ${\mathbf
g}$ on ${\mathbb R}^d$, there exists a unique, one-parameter group
of $C^\infty$ diffeomorphisms $\phi_a^{\,{\mathbf g}}({\mathbf
x}),\,a \in \mathbb{R}$, such that
\begin{equation}
\frac{\partial \phi_a^{\,{\mathbf g}}({\mathbf x})}{\partial a} =
{\mathbf g}(\phi_a^{\,{\mathbf g}}({\mathbf x}))\,,
\end{equation}
with the initial condition $\phi_{a=0}^{\,{\mathbf g}}({\mathbf
x}) \equiv {\mathbf x}$. In a continuous, unitary representation
$U(f)V(\phi)$ of ${\mathcal D}({\mathbb R}^d) \times ${\it
Diff\/}$^c({\mathbb R}^d)$, the local currents are the
self-adjoint generators of $1$-parameter unitary subgroups; so
that
\begin{eqnarray}
U(f) &=& \exp \left[\, \left( i/\mathfrak{m}\right) \rho \left(
f\right) \right] = \exp \left[\, i Q(f,0)\right],\nonumber \\
V( \phi _{s}^{\,\mathbf{g}}) &=& \exp \left[\, \left( is/\hbar
\right) J\left( \mathbf{g} \right) \right] = \exp \left[\, i
Q(0,\mathbf{g})\,\right]. \label{generators}
\end{eqnarray}
The method of induced representations, and other techniques of
unitary group representation, have been extensively used in the
study of the local current algebra.

\section{Some Unitarily Equivalent Representations}

\noindent Let us now turn to some representations of the spatial
components of Eqs.
\negthinspace(\ref{lorentz})-(\ref{covariance}), taken together
with Eqs. \negthinspace(\ref{algebra}). In $d$ space dimensions,
there is a natural representation by derivations over the
(commutative) manifold $\mathbb{R}^{d+1}$, with global coordinates
$(x_{1},x_{2},\dots ,x_{d},w)$. Labeling this representation with
the superscript $(1)$, it is given by \cite{vm2000}
\begin{eqnarray}
q_{j}^{\left(1\right)} &=&i\ell \left( w\frac{\partial }{\partial
x_{j}}
-x_{j}\frac{\partial }{\partial w}\right)\,,  \nonumber \\
M_{jk}^{\left( 1\right) } &=&-i\left( x_{j}\frac{\partial
}{\partial x_{k}}
-x_{k}\frac{\partial }{\partial x_{j}}\right)\,,  \label{rep1} \\
\mathcal{J}^{\left( 1\right) } &=&-i\ell \frac{\partial }{\partial
w}\,,
\nonumber \\
p_{j}^{\left( 1\right) } &=&-i\hbar \frac{\partial }{\partial
x_{j}}\,. \nonumber
\end{eqnarray}
The algebra is represented here by the generators of rigid motions
of $\mathbb{R}^{d+1}$, i.e., the $(d+1)$-dimensional Euclidean
group $E_{d+1}$ which is a semidirect product of the translations
and rotation groups. Note that the coordinate $w$ is like a hidden
dimension in Kaluza-Klein theory \cite{mjd1994}; its presence here
illustrates the idea that sometimes noncommutative structures can
serve as alternatives to hidden dimensions. The Hilbert space on
which the differential operators of Eqs. \negthinspace(\ref{rep1})
act as self-adjoint operators is the space $\mathcal{H} =
\,L^{2}_{d{\mathbf x}dw}$, consisting of complex-valued functions
$\Phi$ on ${\mathbb R}^{d+1}$ that are square-integrable with
respect to the Lebesgue measure $(\,\prod_{j=1}^d dx_j)\,dw$.

But the representation of Eqs. \negthinspace(\ref{rep1}) has the
drawback that when we take the limit as $\,\ell \rightarrow 0$, we
do not recover the usual representation of the Heisenberg algebra
without deformation; rather $q_j^{(1)}$ and $\mathcal{J}^{(1)}$
both tend formally to $0$. We therefore consider an alternative,
denoted with the superscript $(2)$, obtained by introducing a
unitary multiplication operator $U_\ell$ in the Hilbert space
$\mathcal{H}$. Defining $\,U_\ell \,\Phi({\mathbf x}, w) = \exp
[-iw/\ell\,]\,\Phi({\mathbf x}, w)$, we set
\begin{equation}
( q_{j}^{\left( 2\right) },M_{jk}^{\left( 2\right) },\mathcal{J}
^{\left( 2\right) },p_{j}^{\left( 2\right) }) = U_{\ell}\, (
q_{j}^{\left( 1\right) },M_{jk}^{\left( 1\right) },\mathcal{J}
^{\left( 1\right) },p_{j}^{\left( 1\right) }) \,U_{\ell} ^{-1}\,;
\end{equation}
then we have
\begin{eqnarray}
q_{j}^{\left( 2\right) } &=&x_{j}+i\ell \left( w\frac{\partial
}{\partial
x_{j}}-x_{j}\frac{\partial }{\partial w}\right)\,,  \nonumber \\
M_{jk}^{\left( 2\right) } &=&-i\left( x_{j}\frac{\partial
}{\partial x_{k}}
-x_{k}\frac{\partial }{\partial x_{j}}\right)\,,  \label{rep2} \\
\mathcal{J}^{\left( 2\right) } &=&I-i\ell \frac{\partial
}{\partial w}\,,
\nonumber \\
p_{j}^{\left( 2\right) } &=&-i\hbar \frac{\partial }{\partial
x_{j}}\,. \nonumber
\end{eqnarray}
Now the operators smoothly go over to the standard Heisenberg
representation as $\ell \rightarrow 0$. This representation is
also given in \cite{vm2000}.

Let us introduce two corresponding unitarily equivalent
representations obtained by Fourier transformation. As usual, we
have the unitary operator
\[
\mathcal{F}:\,\mathcal{H} = L^{2}_{d{\mathbf x}dw}\,\to
\,\widehat{\mathcal{H}} = L^{2}_{d{\mathbf k}_x dk_w}\,, \]
given by
\[
\left[ \mathcal{F}\Phi\right] \left( k_{1}, \dots, k_{d},
k_{w}\right) = \widehat{\Phi} \left( k_{1}, \dots, k_{d},
k_{w}\right) =\]
\begin{equation} \left( \frac{1}{\sqrt{2\pi
}}\right) ^{d+1}\int_{-\infty }^{\infty }\cdots \int_{-\infty
}^{\infty}\, dx_{1}\cdots dx_d\,dw \,\cdot \label{fourier1}
\end{equation}
\[
\cdot \,\Phi\left( {\mathbf x},w\right) \exp \left[ -i
\mathbf{x}\cdot \mathbf{k}_x-ik_{w}w\,\right],
\]
where ${\mathbf k}_x = (k_{1}, \dots, k_{d})$ and $k_w$ are the
Fourier conjugate variables to ${\mathbf x}$ and $w$ respectively.
Then setting
\begin{eqnarray}
( \widehat{q}_{j}^{\,\left( 1\right) },\widehat{M}_{jk}^{\,\left(
1\right) },\widehat{\mathcal{J}}^{\left( 1\right)
},\widehat{p}_{j}^{\,\left( 1\right) }) &=&
\mathcal{F}\,(q_{j}^{\left( 1\right) },M_{jk}^{\left( 1\right)
},\mathcal{J}^{\left( 1\right)
},p_{j}^{\left( 1\right) }) \,\mathcal{F}^{-1}\,,  \nonumber \\
( \widehat{q}_{j}^{\,\left( 2\right) },\widehat{M}_{jk}^{\,\left(
2\right) },\widehat{\mathcal{J}}^{\left( 2\right)
},\widehat{p}_{j}^{\,\left( 2\right) }) &=& \mathcal{F}\, (
q_{i}^{\left( 2\right) },M_{ij}^{\left( 2\right)
},\mathcal{J}^{\left( 2\right) },p_{i}^{\left( 2\right) })\,
\mathcal{F}^{-1}\,,\nonumber\\  \label{fourier2}
\end{eqnarray}
we have
\begin{eqnarray}
\widehat{q}_{j}^{\,\left( 1\right) } &=& i\ell \left(
k_{w}\frac{\partial }{
\partial k_{j}}-k_{j}\frac{\partial }{\partial k_{w}}\right)\,,
\nonumber \\
\widehat{M}_{jk}^{\,\left( 1\right) } &=& -i\left(
k_{j}\frac{\partial }{
\partial k_{k}}-k_{k}\frac{\partial }{\partial k_{j}}\right)\,,
\label{rep3} \\
\widehat{\mathcal{J}}^{\left( 1\right) } &=&\ell k_{w}\,,  \nonumber \\
\widehat{p}_{j}^{\,\left( 1\right) } &=&\hbar k_{j}\,, \nonumber
\end{eqnarray}
while
\begin{eqnarray}
\widehat{q}_{j}^{\,\left( 2\right) } &=&i\frac{\partial}{\partial
k_j} + i\ell \left( k_{w}\frac{\partial }{
\partial k_{j}}-k_{j}\frac{\partial }{\partial k_{w}}\right)\,,
\nonumber \\
\widehat{M}_{jk}^{\,\left( 2\right) } &=& -i\left(
k_{j}\frac{\partial }{
\partial k_{k}}-k_{k}\frac{\partial }{\partial k_{j}}\right)\,,
\label{rep4} \\
\widehat{\mathcal{J}}^{\left( 2\right) } &=&I + \ell k_{w}\,,
\nonumber \\
\widehat{p}_{j}^{\,\left( 2\right) } &=&\hbar k_{j}\,, \nonumber
\end{eqnarray}
so that in Eqs. \negthinspace(\ref{rep4}), as in Eqs.
\negthinspace(\ref{rep2}), a representation of the Heisenberg
algebra survives in the $\ell \to 0$ limit. Of course the spectra
of the self-adjoint operators representing the generators of the
deformed algebra are the same across all unitarily equivalent
representations.

We focus now for simplicity on the case $d = 1$, which illustrates
well the issues discussed in this article. Because the effective
spatial dimension for the local current algebra will be $d + 1$,
there is also the interesting possibility for $d = 1$ that
features of anyonic statistics of point particles in two-space
could occur. We return to this point in the discussion below.

For $d = 1$, the algebra of Eqs. \negthinspace(\ref{algebra})
reduces to
\begin{eqnarray}
\left[\,q ,p \,\right]
&=& i\hbar \,\mathcal{J}\,, \nonumber \\
\left[\,q ,\mathcal{J}\,\right] &=& -i\frac{\ell ^{2}}{\hbar }
p\,, \label{onedim} \\
\left[\,p , \mathcal{J} \,\right] &=&0\,. \nonumber
\end{eqnarray}
In the representation of Eqs. \negthinspace(\ref{rep3}), it is
useful to introduce polar coordinates $(\rho, \psi)$, with $k_x =
k_{1} = \rho \sin \psi,\, k_{w} = \rho \cos \psi$, and $\,dk_x
dk_w = \rho \,d \rho \,d \psi$. Then the operators become
\begin{eqnarray}
\widehat{q}^{\,\left( 1\right) } &=&i\ell \frac{\partial
}{\partial \psi }\,,
\nonumber \\
\widehat{p}^{\,\left( 1\right) } &=&\hbar \rho \sin \psi\,,
\label{rep3onedim} \\
\widehat{\mathcal{J}}^{\left( 1\right) } &=&\ell \rho \cos \psi\,.
\nonumber
\end{eqnarray}
The representation by Eqs. \negthinspace(\ref{rep3onedim}) in
$\widehat{\mathcal{H}}$ is obviously reducible, since the subspace
of momentum-space wave functions $\widehat{\Phi}$ with support
between $\rho_0$ and $\rho_0 + \Delta \rho$ is invariant under
these operators. Indeed, the Casimir operator
\begin{equation}
C\,=\,\frac{1}{\hbar^{\,2}}\,p^{\,2}\,
+\,\frac{1}{\ell^{\,2}}\,\mathcal{J}^{2}
\end{equation}
commutes with all of the generators in Eqs.
\negthinspace(\ref{onedim}). The corresponding operator
$\widehat{C}$ defined from Eqs. \negthinspace(\ref{rep3onedim})
acts in $\widehat{\mathcal{H}}$ {\it via} multiplication by
$\rho^{\,2}$. The eigenvalues $\,\rho_0^{\,2}\,$ of $C$ label
unitarily inequivalent irreducible representations of
(\ref{onedim}); and we see that the reducible representation given
by Eqs. \negthinspace(\ref{rep3onedim}) acting in
$\widehat{\mathcal{H}}$ is actually a direct integral (from
$\rho_0 = 0$ to $+\infty$) of irreducible representations. The
irreducible component associated with the eigenvalue
$\rho_0^{\,2}$ consists of operators acting on the Hilbert space
$\widehat{\mathcal{H}}_{\rho_0}$ of complex-valued functions
$\widehat{\Phi}_{\rho_0}(\psi)$ on the circle of radius $\rho_0$
centered at the origin in $(k_x,k_w)$-space, that are
square-integrable with respect to the measure $d\psi$.

Let us denote by $\widehat{q}^{\,\left(1\right)}_{\rho_0}$ the
self-adjoint operator $\,i\ell\,\partial/\partial \psi$ acting in
$\widehat{\mathcal{H}}_{\rho_0}$ (defined on a domain of essential
self-adjointness that includes the everywhere essentially
continuous and differentiable functions). Let
$\widehat{p}^{\,\left(1\right)}_{\rho_0}$ and
$\widehat{\mathcal{J}}^{\,\left(1\right)}_{\rho_0}$ be,
respectively, the multiplication operators $\,\hbar \rho_0 \sin
\psi\,$ and $\,\ell \rho_0 \cos \psi$ acting in
$\widehat{\mathcal{H}}_{\rho_0}$. Then a complete orthogonal basis
for $\widehat{\mathcal{H}}_{\rho_0}$ is provided by the
eigenfunctions of $\widehat{q}^{\,\left(1\right)}_{\rho_0}$,
specifically $\,\{\widehat{\Phi}_{\rho_0}^{\,\,(n)}(\psi) =
e^{-in\psi }: n\in {\mathbb Z} \}$, which may be regarded as
infinitely differentiable functions on the circle of radius $\rho
= \rho_0$. Thus $\widehat{q}^{\,\left(1\right)}_{\rho_0}$ has the
eigenvalue spectrum $\{ n\ell: n \in {\mathbb Z}\}$, and the
positional spectrum (which is invariant under unitary equivalence)
is discrete and unbounded in an irreducible self-adjoint
representation of the deformed algebra. On the other hand, the
spectra of $\,\widehat{p}^{\,\left(1\right)}_{\rho_0}\,$ and
$\,\widehat{\mathcal{J}}^{\,\left(1\right)}_{\rho_0}\,$ are
continuous and bounded in absolute value by $\,\hbar \rho_0\,$ and
$\,\ell \rho_0\,$ respectively in such an irreducible
representation.

Note that the complex-valued functions on the plane given by
$\,\widehat{\Phi}^{\,(n)}(\rho, \psi) = e^{-in\psi }, \, n\in
{\mathbb Z}$, do not belong to $\widehat{\mathcal{H}}$ as they are
not square-integrable with respect to $\,\rho \,d\rho \,d \psi$.
These are ``non-normalizable'' eigenfunctions of the operator
$\,\widehat{q}^{\,(1)}\,$ of Eq. \negthinspace(\ref{rep3onedim}).
{\it Bona fide\/} square-integrable eigenfunctions of
$\,\widehat{q}^{\,(1)}\,$ in $\,\widehat{\mathcal{H}}\,$ take the
form $\,f(\rho)\,e^{-in\psi }$, where $\,\int
|f(\rho)|^{\,2}\,\rho\, d\rho\,$ is finite. One regards the
operator $\,\widehat{q}^{\,(1)}\,$ as the direct integral over
$\,\rho\,$ of the measurable field of operators
$\widehat{q}^{\,(1)}_{\rho}$ and the corresponding direct integral
structure of the Hilbert space is given by
\begin{equation}
\widehat{\mathcal{H}} = \int_0^\infty {\widehat{\mathcal{H}}_\rho
} \rho {\kern 1pt} {\kern 1pt} {\rm{d}}\rho.
\end{equation}
Now, in the representations of Eqs. \negthinspace(\ref{onedim})
given by Eqs. \negthinspace(\ref{rep1}) or Eqs.
\negthinspace(\ref{rep2}) acting in $\mathcal{H}$, we have
corresponding decompositions into direct integrals of irreducible
representations over the parameter $\,\rho$ \cite{njv1968}. The
irreducible representations act in Hilbert spaces
$\,\mathcal{H}_{\rho}$, and the operator $\,q^{\,(1)}\,$
[respectively, $\,q^{\,(2)}$] is the direct integral over
$\,\rho\,$ of operators $\,q^{\,(1)}_{\rho}\,$ [respectively,
$\,q^{\,(2)}_{\rho}$] that act in $\,\mathcal{H}_{\rho}$. Let us
introduce polar coordinates $(r,\theta)$, with $\,x=r\sin
\theta\,$ and $\,w=r\cos \theta$. Then the corresponding
eigenfunctions of the operator $q^{\,(1)}_{\rho_0}$, with
eigenvalues $\,n\ell$, in an irreducible representation labeled by
$\,\rho_0$, are the functions $\{\Phi_{\rho_0}^{\,(n)}(r,\theta) =
e^{\,in\theta }: n \in \mathbb{Z}\}$. The corresponding
eigenfunctions of the operator $\,q^{\,(2)}_{\rho_0}\,$ are $\{\,
e^{\,in\theta }e^{-i(r/\ell)\cos \theta }: n \in \mathbb{Z}\}$.
The associated decomposition of $\,\mathcal{H}\,$ as a direct
integral of $\mathcal{H}_{\rho}$ with respect to $\,\rho \,d\rho$,
is developed in \cite{njv1968} using the Fourier-Bessel
transformation. For any element $\,\Phi \left( {x,w} \right)$ of
$\,\mathcal{H}$, we have $\,\Phi(r \sin \theta, r \cos \theta)\,$
= $\,\int_0^{\,\infty}\,\Phi_\rho(r \sin \theta, r \cos
\theta)\rho\,d\rho$, where
\begin{equation}
\Phi _\rho  \left( {r\sin \theta ,r\cos \theta } \right) = (2\pi
)^{-1}\,\int_0^{2\pi } d \theta^{\prime} \int_0^\infty
r^{\prime}dr^{\prime}\times
\end{equation}
\[
\Phi \left( {r^{\prime} \sin \theta^{\prime} ,r^{\prime} \cos
\theta^{\prime} } \right)J_0 \left[\,{\rho \sqrt {r^2  +
(r^{\prime})^2 - 2rr^{\prime} \cos \left( {\theta  -
\theta^{\prime} } \right)} } \,\right].
\]

Finally we can explore the rather subtle way that the usual,
irreducible representation of the Heisenberg algebra must be
recovered from irreducible representations of the deformed algebra
(\ref{onedim}), in the limit $\ell \rightarrow 0$; a point that
was not addressed by Vilela Mendes. With
$[\mathcal{F}\Phi](k_x,k_w) = \widehat{\Phi}(k_x,k_w)$, it is easy
to show that $[\mathcal{F}\left(U_{\ell }\Phi\right)](k_x,k_w) =
\widehat{\Phi}\left( k_{x},k_{w}+1/\ell \right)$. Thus,
considering the representation in $\widehat{\mathcal{H}}$ obtained
from Eqs. \negthinspace (\ref{rep4}) (with $d = 1$), the
irreducible components contributing to the direct integral are
defined from wave functions having support on circles centered at
the point $k_x = 0, k_w = - 1/\ell$ in $(k_x,k_w)$-space; i.e.,
\begin{equation}
k_{x}^{2}+\left( k_{w}+\frac{1}{\ell }\right) ^{2}=\rho _{0}^{2}.
\label{circle}
\end{equation}
To obtain the usual representation of the Heisenberg algebra as
$\ell \rightarrow 0$, we must arrive at wave functions in the
limiting representation that become independent of $k_w$. However,
from Eq. \negthinspace(\ref{circle}), we see that if we try to let
$\ell \rightarrow 0$ while $\rho _{0}$ is held fixed, we obtain no
such limit; rather, $|k_{w}|$ becomes arbitrarily large while the
operator $p^{\,(2)}$ (which acts through multiplication by $\hbar
k_x$) remains bounded. The way out of this difficulty is to allow
$\rho _{0}$ to depend on $\ell $. Taking $\rho _{0}=1/\ell$ in
(\ref{circle}), we have for very small $\ell$ a circle of very
large radius tangent to the horizontal axis at the origin,
approximating the line $k_w = 0$; indeed,
\begin{equation}
k_{w}=-\frac{1}{\ell }+\frac{1}{\ell }\left( 1-k_{x}^{2}\ell
^{2}\right) ^{ \frac{1}{2}}\,;  \label{Lrep}
\end{equation}
so that for any fixed value of $k_x$, $k_w \approx
-(1/2)k_{x}^{2}\ell \to 0$ in the limit as $\ell \rightarrow 0$.
Thus, traversing a parameterized family of irreducible
representations labeled by $\rho _{0}=1/\ell$ (or at least, having
the property that $\rho_0$ tends toward $1/\ell$ as $\ell \to 0$)
is the appropriate way to obtain the Heisenberg algebra in the
limiting irreducible representation.

Moreover, the condition $\rho _{0}=1/\ell$ allows the operator
$p^{\,(2)}$ (whose spectrum is bounded by $\hbar \rho_0$) to
become unbounded as desired when $\ell \to 0$.

\section{The Kinetic Energy and Harmonic Oscillator Hamiltonians}

\noindent To investigate in detail the quantum-mechanical behavior
of a particle system described by the deformed algebra, such as
the harmonic oscillator with a fundamental length scale, we need
to settle on the form of the kinetic energy part of the
Hamiltonian $H_0$, and write the total Hamiltonian $H = H_0 + V$.
In \cite{vm2000} it was suggested that for a particle of mass
$\mathfrak{m}$, we should use $H_{0}= p^{\,2}/2\mathfrak{m}$,
where $p$ is the generator appearing in the algebra of Eqs.
\negthinspace(\ref{onedim}), and that the oscillator Hamiltonian
should then be $H_{\mathrm {osc}} \,=\, p^{\,2}/2\mathfrak{m}
\,+\, \mathfrak{m}\,\omega^{\,2} q^{\,2}/2$.

But one reasonable criterion for determining the choice of $H_0$
is the physical condition that the time-derivative of the particle
position should be the particle velocity. That is, we should
expect $H_0$ and $H_{\mathrm{osc}}$ to satisfy,
\begin{equation}
\dot{q}\,=\,\frac{1}{i\hbar }\left[ \,q,H_0\right] =
\frac{1}{i\hbar }\left[ \,q,H_{\mathrm{osc}}\right] =
\frac{p}{\mathfrak{m}}\,. \label{criterion}
\end{equation}
However, we have from Eqs. \negthinspace(\ref{onedim}) that
\begin{equation}
\frac{1}{i\hbar }\left[ \,
q,\frac{p^{\,2}}{2\mathfrak{m}}\,\right] =
\frac{1}{2\mathfrak{m}}\left(
p\,\mathcal{J}+\mathcal{J}p\,\right)\,,
\end{equation}
which becomes $p/\mathfrak{m}$ when $\mathcal{J}$ is the identity
operator, but not otherwise. To fulfill Eq. (\ref{criterion}) we
propose to modify the form of the kinetic energy term in the
Hamiltonian, so that
\begin{equation}
H_0=\frac{1}{2\mathfrak{m}}\left\{ p^{\,2}+\frac{\hbar
^{\,2}}{\ell ^{\,2}}\left( \mathcal{J}- I \right) ^{2}\right\}\,,
\label{choice}
\end{equation}
where $I$ is the identity operator in a representation of the
algebra. Note that the coefficient ${\hbar ^{\,2}}/{\ell ^{\,2}}$
in (\ref{choice}) is needed to obtain the correct commutation
relation with $q$.

Now, in the representation $q^{(2)},\,p^{(2)},\,\mathcal{J}^{(2)}$
of the deformed algebra given by Eqs. \negthinspace(\ref{rep2}),
with $d = 1$, we have
\begin{equation}
H_0^{(2)}=-\frac{\hbar ^{2}}{2\mathfrak{m}}\left\{ \frac{\partial
^{2}}{\partial x^{2}}+\frac{%
\partial ^{2}}{\partial w^{2}}\right\}\,,  \label{choice1}
\end{equation}
while in the Fourier transformed representation
$\widehat{q}^{(2)},\,\widehat{p}^{(2)},\,\widehat{\mathcal{J}}^{(2)}$,
we have
\begin{equation}
\widehat{H}_0^{(2)} =  \frac{\hbar ^{2}}{2\mathfrak{m}} (k_x^2 +
k_w^2)\,. \label{hamkspace}
\end{equation}
The explicit dependence of $H_0$ on $\ell$ has disappeared.

Nevertheless, this is the family of irreducible representations
that goes over smoothly to the standard representations of the
Heisenberg algebra as we take the limit $\ell \to 0$, with $\rho_0
= 1/\ell$. Returning to the discussion following Eq.
\negthinspace(\ref{Lrep}), we have from Eq.
\negthinspace(\ref{hamkspace}),
\begin{equation}
\widehat{H}_0^{(2)} \,\approx\,  (\hbar ^{2}/2\mathfrak{m}) [k_x^2
+ (\ell^2/4) k_x^4]\,,
\end{equation}
so that $\,\widehat{H}_0^{(2)} \to  (\hbar ^{2}/2\mathfrak{m})
k_x^2\,$ as $\ell \to 0$.

Now the oscillator potential, and other potentials of the form
$V(q)$, commute with $q$; so that $[\,q, H_0 + V] = [\,q,H_0]$,
and it is appropriate to take the harmonic oscillator Hamiltonian
for deformed quantum mechanics to be $H_{\mathrm{osc}} = H_0 +
\mathfrak{m}\,\omega^{\,2} q^{\,2}/2$, with $H_0$ as in Eq.
\negthinspace(\ref{choice}). From (\ref{rep2}), however, we see
that the potential energy term no longer acts {\it via\/}
multiplication in $L^{\,2}_{dxdw}$, but as a differential operator
{\it via\/} further derivative terms.

\section{Discretized and Extended Local Current Algebras}

\noindent In this section we explore two approaches to the
introduction of local currents.

In the first approach, we interpret locality with respect to a
basis of eigenvectors of $q$ in an irreducible representation of
the deformed Heisenberg algebra (\ref{onedim}). This leads to
discretized mass and momentum density operators; i.e., a theory on
a lattice.

In the second approach, we interpret ``locality'' with respect to
the $(xw)$-space on which representations of (\ref{onedim}) are
modeled. Here we consider two possibilities. The first is to work
straightforwardly with the nonrelativistic current algebra in
two-dimensional space (LCA2) to describe the one-dimensional,
deformed quantum kinematics, while the second is to introduce a
semidirect sum of LCA2 with the algebra of vector fields on the
line. In the discussion, we note the tension occurring between
locality and irreducibility.

\subsection{Locality with respect to the discrete \\ positional
spectrum}

In an irreducible representation of Eqs.
\negthinspace(\ref{rep3onedim}), the spectrum of the self-adjoint
operator representing the generator $q$ is discrete, given by
$n\ell$ for $n \in \mathbb{Z}$; write the corresponding
eigenvector as $\left| n\ell \right\rangle $. We shall then write
$\,q\,|n\ell\rangle = n\ell|n\ell\rangle$, and
\begin{equation}
q = \sum_{n=-\infty}^{\infty} n\ell|n\ell\rangle \langle
n\ell|\,,\quad I = \sum_{n=-\infty}^{\infty} |n\ell\rangle \langle
n\ell|\,. \label{discrete0}
\end{equation}
The corresponding local mass density operator
$\,\mathfrak{J}_{\,q}\,$ takes the form
\begin{equation}
\mathfrak{J}_{\,q}(g) = \mathfrak{m}\sum_{n\,=-\infty }^{\infty
}g(n\ell) \left| n\ell \right\rangle \left\langle n\ell \right|\,,
\label{discrete1}
\end{equation}
where in analogy with the continuum case, the real-valued function
$g$ has compact support; i.e., for some $N > 0$, $g(n\ell) = 0$
whenever $|n\ell| > N\ell$. When $g(n\ell)$ approximates the
function $\,n\ell$, $\mathfrak{J}_{\,q}(g)$ approximates the
moment operator $\,\mathfrak{m}q$. When $g(n \ell) \geq
0\,\,(\forall n \in \mathbb{Z})$, $\mathfrak{J}_{\,q}(g)$ is a
positive operator. When $g(n\ell)$ approximates the constant
function $1$, $\mathfrak{J}_{\,q}(g)$ approximates the mass times
the identity operator. As we are in a representation of the local
currents describing a single particle, we can interpret
$(1/\mathfrak{m})\mathfrak{J}_{\,q}(g)$ as a spatial probability
density operator averaged with $g(n\ell)$.

Eqn. \negthinspace(\ref{rep3onedim}) implies
\begin{equation}
\left\langle n\ell \right| p\left| m\ell \right\rangle
=\frac{\hbar
\rho _{0}%
}{2i}\left( \delta _{n+1,m}-\delta _{n-1,m}\right)
\label{matrixelt1}
\end{equation}
and
\begin{equation}
\left\langle n\ell \right| \mathcal{J}\left| m\ell \right\rangle
=\frac{%
\ell \rho _{0}}{2}\left( \delta _{n+1,m}+\delta _{n-1,m}\right).
\label{matrixelt2}
\end{equation}
Introduce the local currents
\begin{equation}
\mathfrak{J}_{\,p}(h) = \frac{1}{2}\sum_{n=-\infty }^{\infty }
\widetilde{h}( n\ell) \{\, p \left| n\ell \right\rangle
\left\langle n\ell \right| +\left| n\ell \right\rangle
\left\langle n\ell \right| p\,\}  \label{discrete2}
\end{equation}
and
\begin{equation} \mathfrak{J}_{\,\mathcal{J}}\left( r\right)
=\frac{1}{2}\sum_{n=-\infty }^{\infty }\widetilde{r}\left( n\ell
\right) \{\, \mathcal{J}\left| n\ell \right\rangle \left\langle
n\ell \right| +\left| n\ell \right\rangle \left\langle n\ell
\right| \mathcal{J}\,\}, \label{discrete3}
\end{equation}
where $h\left( n\ell \right) \equiv (1/2)\,[\,\widetilde{h}\left(
n\ell \right) + \widetilde{h}\left( \left( n+1\right) \ell
\right)\,]$ and $r\left( n\ell \right) \equiv
(1/2)\,[\,\widetilde{r}\left( n\ell \right) +\widetilde{r}\left(
\left( n+1\right) \ell \right)\,]$ are also taken to be compactly
supported. As $\,\widetilde{h}(n\ell)\,$ and
$\,\widetilde{r}(n\ell)\,$ approximate the function that is
identically $1$, so do $\,h(n\ell)\,$ and $\,r(n\ell)$; then
$\mathfrak{J}_{\,p}(h)$ approximates $\,p$,
$\mathfrak{J}_{\,\mathcal{J}}\left( r\right)$ approximates
$\,\mathcal{J}$, and the global algebra is recovered.

From Eqs. \negthinspace(\ref{matrixelt1})-(\ref{discrete3}), we
find
\[
\mathfrak{J}_{\,p}\left( h\right) \,=\,\frac{\hbar \rho
_{0}}{2i}\sum_{n=-\infty }^{\infty }h\left( n\ell \right) \{\,
\left| n\ell \right\rangle \langle \left( n+1\right) \ell |
\nonumber
\]
\begin{equation}
-\left| \left( n+1\right) \ell \right\rangle \left\langle n\ell
\right| \,\}\,,  \label{discrete4}
\end{equation}
\[
\mathfrak{J}_{\,\mathcal{J}}\left( r\right) \,=\,\frac{\ell \rho
_{0}}{2} \sum_{n=-\infty }^{\infty }r\left( n\ell \right) \{\,
\left| n\ell \right\rangle \left\langle \left( n+1\right) \ell
\right| \nonumber
\]
\begin{equation}
+\left| \left( n+1\right) \ell \right\rangle \left\langle n\ell
\right| \,\}\,. \label{discrete5}
\end{equation}
For the Lie algebra of currents generated by these operators, in
the irreducible representation labeled by $\rho _{0}$, to be {\it
local,\/} we need the commutator brackets of the operators
$\,\mathfrak{J}_{\,q}\left( g\right)$, $\mathfrak{J}_{\,p}\left(
h\right)$, and $\mathfrak{J}_{\,\mathcal{J}}\left( r\right)$ given
by Eqs. \negthinspace(\ref{discrete1}), (\ref{discrete4}), and
(\ref{discrete5}) to yield similarly local expressions. These
expressions are all linear combinations of operators of the form
$\,\left| n\ell \right\rangle \left\langle n\ell \right|$,
$\,\left| n\ell \right\rangle \left\langle \left( n+1\right) \ell
\right|$, and $\,\left| \left( n+1\right) \ell \right\rangle
\left\langle n\ell \right|$.

In fact, we have
\begin{equation}
[ \mathfrak{J}_{\,q}\left( g_{1}\right) ,\mathfrak{J}_{\,q}\left(
g_{2}\right)] =0\,,
\end{equation}
\begin{equation}
\left[ \mathfrak{J}_{\,q}\left( g\right) ,\mathfrak{J}_{\,p}\left(
h\right) \right] =-i\frac{\mathfrak{m}\hbar }{\ell
}\mathfrak{J}_{\,\mathcal{J}}\left({r} \right)\,,
\end{equation}
where ${r}\left( n\ell \right) = h(n\ell)\,\{\, g\left( n\ell
\right) -g\left( \left[ n+1\right] \ell \right) \}$, and
\begin{equation}
\left[ \mathfrak{J}_{\,q}\left( g\right)
,\mathfrak{J}_{\,\mathcal{J}}\left( r\right) \right]
=i\frac{\mathfrak{m}\ell }{\hbar }\mathfrak{J}_{\,p}\left({h}
\right)\,,
\end{equation}
where ${h}\left( n\ell \right) = r\left( n\ell \right) \left(
g\left( n\ell \right) -g\left( \left[ n+1\right] \ell \right)
\right)$, which thus far are satisfactorily local. But other
commutators, such as $\,\left[ \mathfrak{J}_{\,p}\left( h\right)
,\mathfrak{J}_{\,\mathcal{J}}\left( r\right)\right]$, generate
terms of the form  $\left| \left( n+1\right) \ell \right\rangle
\left\langle \left( n-1\right) \ell \right|$ and $\left| \left(
n-1\right) \ell \right\rangle \left\langle \left( n+1\right) \ell
\right| $, so that successive commutators generate additional
terms $\,\left| \left( n-m\right) \ell \right\rangle \left\langle
\left( n+m\right) \ell \right|\,$ and $\,\left| \left( n+m\right)
\ell \right\rangle \left\langle \left( n-m\right) \ell \right|$,
for arbitrary $m \in \mathbb{Z}$. Therefore, to close the Lie
algebra of these currents, one is forced to include new basis
elements in the (already infinite-dimensional) current algebra,
having more general forms; e.g.,
\[
\sum_{n,m = -\infty}^\infty s(n\ell,m\ell)\{\,\left| \left(
n+m\right) \ell \right\rangle \left\langle \left( n-m\right) \ell
\right|
\]
\[ \pm \left| \left( n-m\right) \ell \right\rangle \left\langle \left(
n+m\right) \ell \right|\,\}\,, \]
where $s$ is a compactly supported function on the square lattice
of points $(n\ell,m\ell)$. Such currents are {\it nonlocal\/} in
the positional eigenvalues, since $(n-m)\ell$ and $(n+m)\ell$
become arbitrarily far apart. This sort of behavior by the
commutation relations of discretized local derivatives is
well-known in the context of lattice models.

Before leaving the discussion of the discretized current algebra,
it is worth remarking that we do have within this framework an
{\it equation of continuity\/} for the deformed quantum theory,
relating the time-derivative of $\frak{J}_{\,q}$ to the spatial
divergence of $\frak{J}_{p}$. Taking the Hamiltonian $H$ to be
$H_0 + V(q)$, with $H_0$ given by Eq. \negthinspace(\ref{choice}),
we have
\[
\dot{\frak{J}}_{\,q}(g)\,=\, \frac{1}{i\hbar}\left[\,
\frak{J}_{\,q}(g), H \,\right] \,=\,
\frac{\mathfrak{m}}{i\hbar}\left[\, \sum_{n=-\infty}^\infty
g\left( n\ell\right) \left| n\ell\right\rangle \left\langle
n\ell\right| , H\,\right] \]
\begin{equation} =
\frac{\mathfrak{m}}{i\hbar}\left[\,\sum_{n=-\infty}^\infty g\left(
n\ell\right) \left| n\ell\right\rangle \left\langle n\ell\right| ,
H_0\right]\,. \label{begincontinuity}
\end{equation}
Straightforward calculations yield
\[
\left[ \,\sum_{n}g\left(  n\ell\right)  \left|  n\ell\right\rangle
\left\langle n\ell\right|,p^{\,2}  \,\right]\]
\[
=\left(  \frac{\hbar\rho_{0}}{2i}\right)  ^{2}
\sum_{n}g\left( n\ell\right)
\left\{  -\left|  (n+2)\ell\right\rangle \left\langle n\ell\right|
+\left| n\ell\right\rangle \left\langle (n+2)\ell\right|  \right\}
\]
\[
+ \left(  \frac{\hbar\rho_{0}}{2i}\right)  ^{2}
\sum_{n}g\left( n\ell\right)
\left\{  -\left| (n-2)\ell\right\rangle \left\langle n\ell\right|
+\left| n\ell\right\rangle \left\langle (n-2)\ell\right|  \right\}
\]
\[
=\left( \frac{\hbar\rho_{0}}{2i}\right) ^{2} \sum_{n}\left(
g\left( (n-2)\ell\right)  -g\left(  n\ell\right)  \right)\,\cdot
\]

\begin{equation}
\cdot \,\left\{ \,\left| (n-2)\ell\right\rangle \left\langle
n\ell\right| -\left| n\ell\right\rangle \left\langle
(n-2)\ell\right| \right\},
\end{equation}
while
\[
\left[ \,\sum_{n}g\left(  n\ell\right) \left| n\ell\right\rangle
\left\langle n\ell\right|,\left(  \mathcal{J}-I\right)  ^{2} \,
\right] =
\]
\[
-\frac{\ell^{2}\rho_{0}^{2}}{4}\,\sum_{n}g\left( n\ell\right)
\left\{\, \left|  (n-2)\ell\right\rangle \left\langle n\ell\right|
-\left| n\ell\right\rangle \left\langle (n-2)\ell\right|\right\}
\]
\[
- \frac{\ell^{2}\rho_{0}^{2}}{4}\,\sum_{n}g\left( n\ell\right)
\left\{\,\left| (n+2)\ell\right\rangle \left\langle n\ell\right|
-\left| n\ell\right\rangle \left\langle (n+2)\ell\right|  \right\}
\]
\[
+\ell\rho_{0}\,\sum_{n}g\left( n\ell\right) \cdot \left\{\,\left|
(n-1)\ell\right\rangle \left\langle n\ell\right| -\left|
n\ell\right\rangle \left\langle (n-1)\ell\right|\,\right\}
\]
\[
+\ell\rho_{0}\,\sum_{n}g\left( n\ell\right) \cdot \left\{\,\left|
(n+1)\ell\right\rangle \left\langle n\ell\right| -\left|
n\ell\right\rangle \left\langle (n+1)\ell\right| \right\}
\]
\[
=  -\frac{\ell^{2}\rho_{0}^{2}}{4} \, \sum _{n} \left( g\left(
n\ell\right) -g\left( (n-2)\ell\right) \right)\cdot \]
\[\cdot \left\{ -\left| n\ell\right\rangle \left\langle
(n-2)\ell\right| +\left| (n-2)\ell\right\rangle \left\langle
n\ell\right| \right\}
\]
\[
+\ell\rho_{0} \, \sum _{n}\left( g\left( n\ell\right) -g\left(
(n-1)\ell\right)  \right)\cdot
\]
\[
\cdot \left\{ \, -\left|  n\ell\right\rangle \left\langle
(n-1)\ell\right| +\left| (n-1)\ell\right\rangle \left\langle
n\ell\right| \,\right\}.
\]
Then from Eq. \negthinspace(\ref{begincontinuity}), with $H_0$ as
in Eq. \negthinspace(\ref{choice}), we obtain (after replacing the
index $n$ by $n+1$ in the infinite sum)
\[
\dot{\mathfrak{J}}_{\,q}(g)\,=\, \frac{
\hbar\rho_{0}}{2i}\,\sum_{n}\, \frac{\left( g(\left(
n+1\right)\ell) -g\left(  n\ell\right) \right) }{\ell}\,\cdot
\]
\[
\cdot \left\{ \,-\left| (n+1)\ell\right\rangle \left\langle
n\ell\right| +\left| n\ell\right\rangle \left\langle
(n+1)\ell\right| \right\}
\]
\begin{equation}
= \mathfrak{J}_{p}\left(  Dg\right)\,, \label{discretecontineq}
\end{equation}
where \begin{equation} Dg\left( n\ell\right) \equiv\frac{g\left(
(n+1)\ell\right) -g\left( n\ell\right) }{\ell}\,
\end{equation}
is the discretized derivative. Evidently Eq.
\negthinspace(\ref{discretecontineq}) is precisely the required
continuity equation. The density ${\frak J}_{\,q}$ and current
${\frak J}_{\,p}$ that appear in this equation of continuity are
local, but they belong to a Lie algebra that necessarily includes
currents that are nonlocal with respect to the positional operator
$q$.

\subsection{Locality with respect to the extended \\ spatial manifold}

Consider again the global Lie algebra given by Eqs.
\negthinspace(\ref{onedim}). A quite different approach to
introducing local currents is suggested by the form of the
representation of this Lie algebra by Eqs.
\negthinspace\negthinspace(\ref{rep2}), with $d = 1$. The idea is
to define a current algebra that is local in $(x,w)$-space, from
which---with the right choices of limiting test functions---we
shall be able to recover Eqs. \negthinspace(\ref{rep2}).

Thus let us refer back to the LCA of Eqs.
\negthinspace(\ref{semidirectsum1}), and interpret these equations
as applying in a two-dimensional Euclidean space with coordinates
$(x,w)$, extending the spatial manifold by one dimension. We then
have the operator-valued distribution $Q\left( h,g_{x},g_{w}
\right)$ acting in $L^{2}_{dxdw}$, where $h$ is drawn from the
space of compactly-supported, real-valued $C^{\infty}$ test
functions on $(x,w)$-space, and $g_{x},g_{w}$ are the components
of a compactly-supported, $C^\infty$ vector field on
$(x,w)$-space:
\begin{eqnarray}
Q( h,g_{x},g_{w}) = h\left( x,w\right) \,\,+& & \nonumber\\
+\,\frac{1}{2i}\,\{\, g_{x}\left( x,w\right) \frac{\partial
}{\partial x} &+& \frac{\partial }{
\partial x}\,g_{x}\left( x,w\right) \} \nonumber \\
+\,\frac{1}{2i}\,\{\,g_{w}\left( x,w\right)
\frac{\partial}{\partial w} &+& \frac{\partial }{\partial w}
\,g_{w}\left( x,w\right) \}\,. \label{lca2}
\end{eqnarray}
Defining $Q_\ell ( h,g_x,g_w) = Q (h, \ell g_x, \ell g_w)$ for
$\ell > 0$, we obtain a family of operators parameterized by
$\ell$. In the $\ell \to 0$ limit, $Q_\ell( h,g_x,g_w)$ reduces to
$Q(h,0,0)$. Then with
\begin{equation}
\rho(f) \,=\, \lim_{h \to f} \mathfrak{m}\,Q(h,0,0)\,,
\label{rholim}
\end{equation}
we recover the earlier one-particle mass density operator in one
space dimension. The limit here pertains to the fact that $f$
depends only on $x$ and is independent of $w$, while $h$ is
compactly supported in $(x,w)$-space.

The local current $Q_\ell ( h,g_x,g_w)$ is clearly motivated by
the form of $q^{\left( 2\right) }$ in Eq.
\negthinspace(\ref{rep2}); in fact, in the limit where $h(x,w)$
approaches the coordinate function $x$, and the vector field
$(g_x(x,w),g_w(x,w))$ approaches $(-w,x)$, we recover $q^{(2)}$
with the space dimension $d = 1$. Evidently, $Q$ (or,
alternatively, $Q_\ell$) is also sufficiently general to let us
obtain the other global operators in the deformed current algebra,
when suitable limits of $h,g_{x}$ and $g_{w}$ are taken. Thus the
operator $p^{(2)}$ (with $d = 1$) is just $Q(0,\hbar g_x,0)$ or
$Q_\ell(0, (\hbar/\ell)g_x,0)$, taken in the limit where $g_x$
approaches the constant vector field of magnitude $1$. Likewise
$\mathcal{J}^{(2)}$ is $Q(h,0,\ell g_w)$ or $Q_\ell(h,0,g_w)$,
taken in the limit where both $h$ and $g_w$ become identically
$1$.

The natural choices of local currents corresponding to $p^{(2)}$
and $\mathcal{J}^{(2)}$ are, respectively,
\begin{equation}
J(g)\,=\,\frac{\hbar }{2i}\,\{\,g\left( x\right) \frac{\partial
}{\partial x}+%
\frac{\partial }{\partial x}g\left( x\right)\} \label{submaster1}
\end{equation}
and
\begin{equation}
\mathcal{J}(k) =k\left( w\right) +\frac{\ell }{2i}\,\{ k\left(
w\right) \frac{\partial }{\partial w}+\frac{\partial }{\partial
w}%
k\left( w\right) \}\,,  \label{submaster2}
\end{equation}
where $g(x)$ and $k(w)$ are compactly-supported $C^{\,\infty}$
functions on $\mathbb{R}$. These local currents incorporate the
intuitive idea of local flows in the two coordinate directions. To
express them in terms of $Q$ or $Q_\ell$, we write (again
recalling that the arguments of $Q$ are compactly supported in
both the $x$ and $w$ coordinates):
\begin{equation}
J(g) = \lim_{g_x \to g} Q(0,\hbar g_x, 0) = \lim_{g_x \to g}
Q_\ell(0,\frac{\hbar}{\ell}\, g_x, 0) \label{submaster3}
\end{equation}
and
\begin{equation}
\mathcal{J}(k) = \lim_{h \to k} \lim_{g_w \to k} Q(h,0,\ell g_w) =
\lim_{h \to k} \lim_{g_w \to k} Q_\ell(h,0,g_w)\,.
\label{submaster4}
\end{equation}
Because of the way the operator $q^{(2)}$ mixes the $x$ and the
$w$ directions, it is necessary to incorporate the full
$(x,w)$-dependence in the test functions $h,g_{x}$ and $g_{w}$
that appear as arguments of $Q$. Then the current algebra that
accommodates all the natural local and global limits is just LCA2;
i.e., the algebra of the $Q(h, g_x, g_w)$ satisfying the
semidirect sum Lie algebra of Eq.
\negthinspace(\ref{semidirectsum2}) with $d = 2$. So we have the
{\it usual\/} local current algebra of nonrelativistic quantum
mechanics, but localized in two space dimensions rather than just
one.

An interesting feature of this framework is that the decomposition
of $L^2_{dxdw}(\mathbb{R}^2)$ into a direct integral of
irreducible representations of the global algebra (labeled, as in
Sec. \negthinspace III, by $\rho_0$), is not respected by the
one-particle irreducible representation of the LCA. That is, the
{\it local\/} currents unavoidably {\it connect\/} the reducing
subspaces of the global algebra. We have a kind of tension between
our desire to incorporate local currents, and the assumption that
we can work with a single, irreducible representation of the
global, deformed algebra.

This situation does not occur for the usual Heisenberg algebra,
where the Hilbert space for a single irreducible representation
labeled by $\hbar$ also carries the one-particle representation of
the full LCA. However, it is reminiscent of earlier results
pertaining to self-adjoint representations of the LCA describing
{\it spinning\/} particles in three space dimensions. Here
irreducible representations of the global algebra describe quantum
particles with fixed spin---i.e., the operators act within a
single irreducible representation of $SU(2)$---while the local
currents inevitably contain spin-changing terms, that connect
representations associated with a tower of different spins
\cite{gag1983c}.

Let us take another look at how we can recover LCA1 as the $\ell
\to 0$ limit of LCA2 in the single-particle representation written
above. If we take the operators $Q_\ell$ as our starting point,
with fixed test functions $(h,g_x,g_w)$ independent of $\ell$,
then as $\ell \to 0$ and $h \to f$, we recover only the density
operator $\rho(f)$, not the full LCA1. To recover the local
currents $J(g)$ in the $\ell \to 0$ limit, we must allow at least
some of the test functions themselves to be $\ell$-dependent, as
in Eq. \negthinspace(\ref{submaster3}). But the form of $J(g)$ in
Eq. \negthinspace(\ref{submaster1}) is actually independent of
$\ell$. This suggests that, for given $\ell$, we consider an {\it
extension\/} of the current algebra LCA2 (generated by the
$Q_\ell(h,g_x,g_w)$) by the algebra of vector fields on the line
(generated by the $J(g)$ in Eq. \negthinspace(\ref{submaster1})),
via the bracket
\begin{equation}
\left[\, Q_\ell\left(h,g_{x},g_{w}\right) ,J\left( g\right)
\right] =i\hbar Q_\ell (
\widetilde{h},\widetilde{g}_x,\widetilde{g}_{w}),
\label{extension3}
\end{equation}
where
\begin{eqnarray}
\widetilde{h}\left( x,w\right) &=& g\left( x\right) \frac{
\partial }{\partial x}\,h\left( x,w\right) , \nonumber \\
\widetilde{g}_x\left( x,w\right) &=& g\left( x\right)
\frac{\partial }{\partial x}\,g_x\left( x,w\right) -g_x\left(
x,w\right)\frac{\partial }{\partial x}\,g\left(
x\right)\,,\nonumber \\
\widetilde{g}_w\left( x,w\right) &=& g\left( x\right) \frac{
\partial }{\partial x}\,g_w (x,w)\,.
\label{extension4}
\end{eqnarray}
Note in Eqs. \negthinspace(\ref{extension3})-(\ref{extension4})
that $h, \widetilde{h}$ are compactly-supported, $C^\infty$
functions on $(x,w)$-space, $(g_x, g_w)$ and $(\widetilde{g_x},
\widetilde{g_w})$ are compactly-supported, $C^\infty$ vector
fields on $(x,w)$-space, while $g$ is a compactly-supported,
$C^\infty$ vector field in the $x$-coordinate only.

Now in the limit $\ell \to 0$, $Q_\ell(h,g_x,g_w)$ becomes
multiplication by $h(x,w)$, while $J(g)$ survives as the operator
for total momentum density in the $x$-direction, independent of
(or integrated over) $w$. The representation is still reducible;
$w$ has become a kind of unobservable, internal coordinate for a
particle theory in one space dimension. This construction also
generalizes to higher space dimensions, augmented by the one
additional coordinate $w$. Note, however, that the form of the
kinetic energy term in the Hamiltonian, given by $H_0^{(2)}$ in
Eq. \negthinspace(\ref{choice1}), is {\it independent\/} of
$\ell$; the second derivative with respect to $w$ does not vanish
as $\ell \to 0$. Upon taking this limit, we can use the fact that
the operators $\rho(f)$ and $J(g)$ commute with
$\partial^{2}/\partial w^2$ to recover the continuity equation in
the continuum,
\begin{equation}
\dot{\rho}(f)\,=\, \frac{1}{i\hbar}\left[\, \rho (f), H_0
\,\right] \,=\,J(\frac{d f}{d x}), \end{equation}
and the ordinary quantum mechanics of a free particle having $x$
as its positional coordinate.

For local currents as defined by Eq. \negthinspace(\ref{lca2}) in
$(x,w)$-space, we can also write an equation of continuity for a
free particle,
\begin{equation}
\frak{m}\,\dot{Q}(h,0,0) = \frac{\frak{m}}{i\hbar}\left[\, Q
(h,0,0), H_0 \,\right] = \hbar Q(0, \frac{\partial h}{\partial x},
\frac{\partial h}{\partial w}).\quad
\end{equation}
But note that $Q(h,0,0)$ is no longer a {\it positional\/} mass
density. If the potential energy $V$ is a function of the
(deformed) position operator $q$ (as in the case of the deformed
harmonic oscillator), then $V(q)$ does not commute with
$\frak{m}\,\dot{Q}(h,0,0)$ and we have no such continuity
equation.

\section{Concluding Remarks}

\noindent In the case of one space dimension, we have described
two approaches to introducing fixed-time local currents for a
subalgebra of the deformed Poincar\'e-Heisenberg algebra discussed
by Vilela Mendes. The first requires a nonlocal Lie algebra
generated by discretized local currents, but the currents act
within an irreducible representation of the global algebra. The
second requires adjoining an extra dimension to the spatial
manifold, and the local currents connect the reducing subspaces in
a direct integral of irreducible representations of the global
algebra.

From our perspective, the latter approach offers some attractive
possibilities. We have mentioned above the existence of many
interesting, inequivalent representations of LCA2, including
representations describing $N$ particles satisfying the statistics
of anyons. This suggests a possible new interpretation of such
representations---not as describing conventional particles in
two-space, but as describing local currents for a deformed algebra
of quantum mechanics having some anyonic properties. Such an
interpretation is a topic of continuing investigation by the
authors.

\section{Acknowledgment}

\noindent G. Goldin wishes to thank the Leverhulme Trust for
support during the course of this research, and King's College
London for hospitality during his 2004-2005 sabbatical leave.

\end{document}